\documentclass{article}

\usepackage{amsmath, amssymb, amsthm}
\usepackage{booktabs}
\usepackage{xcolor}

\usepackage{hyperref}

\usepackage{orcidlink}

\newtheorem{Proposition}{Proposition}
\newtheorem{Corollary}{Corollary}
\newtheorem{Theorem}{Theorem}
\newtheorem*{Proof}{Proof}

\newtheorem{Remark}{Remark}
\newtheorem{Example}{Example}
\newtheorem{Algorithm}{Algorithm}

\newcommand{\dcov}{d^{\nabla}}

\begin{document}

\begin{center}
{\LARGE\bfseries Decomposition and characterization of curl forces for all space dimensions}\\
{\large Rados\l aw Antoni Kycia$^{1,a}$} \orcidlink{https://orcid.org/0000-0002-6390-4627} \\
\small
$^{1}$ PK Krakow University of Technology, Faculty of Computer Science and Mathematics\\
Warszawska 24, Krak\'ow, 31-155, Poland\\
$^{a}$\texttt{kycia.radoslaw@gmail.com}
\end{center}

\begin{abstract}
\noindent
We present a local decomposition of classical force fields in arbitrary dimension into gradient and curl parts by representing a force as a differential $1$-form (the work form). Using the radial homotopy operator on a star-shaped domain centered at $x_{0}$, we split the work form into an exact (gradient) component and an antiexact component, which generalizes the three-dimensional curl force. We prove that applying the Frobenius decomposition to the antiexact part with the same center $x_{0}$ vanishes, and we resolve this problem by shifting the homotopy center to a point $x_{1}\neq x_{0}$ where the antiexact form does not vanish. This separates the antiexact component into an integrable part determined by generalized potentials and a non-integrable remainder. Unlike the Darboux classification, the general and recursive decompositions avoid solving partial differential equations, which arise only when seeking a closed connection form for the gradient-recursive case. We also describe a single-center variant, establish a three-dimensional class-reduction property for homogeneous curls, analyze the coalescing-centers limit $x_{1}\rightarrow x_{0}$ and its renormalization, and discuss the local-to-global extension. Finally, we show that the conservative component is invariant under spatial amplitude modulation of the antiexact part. By analyzing the resulting conformal class of curl forces reveals a formal Weyl-geometric structure on a trivial real line bundle that unifies passive gauge shifts with active Weyl symmetries and admits natural mechanical and Carath\'{e}odory thermodynamical interpretations.
\end{abstract}

\noindent
\textbf{Keywords:} curl force; nonconservative forces; antiexact forms; homotopy operator; geometric decomposition; Frobenius theorem; Weyl geometry\\
\textbf{Mathematical Subject Classification:} 58A10, 53Z05, 51P05.

\section{Introduction}
A well-known result in classical mechanics \cite{Thirring, AbrahamMarsden, Frankel} is that a force with a vanishing curl is the gradient of a scalar function known as a potential. On a Riemannian manifold \cite{BennTucker, Bleecker, deRham, Thirring, KobayashiNomizu, Sternber, AbrahamMarsden, Frankel, LoringTu} $M$ with a metric tensor $g:TM\times TM\rightarrow \mathbb{R}$, a force can be represented by a differential $1$-form (the work form). In this framework, the existence of a potential is equivalent to the statement that the work form is exact, meaning it is the exterior derivative of a $0$-form (a scalar function), which is the potential. This property holds locally by the Poincar\'{e} lemma \cite{EdelenExteriorCalculus, EdelenIsovectorMethods, ExteriorSystems}.

A force $F$ satisfying the condition $\nabla \times F \neq 0$ is known as a curl force, also referred to as a pseudo-gyroscopic or circulatory force. Such a force can still be decomposed into gradient-like components, although this decomposition is not unique. These forces are of physical significance \cite{CurlForces1, CurlForces2, CurlForces3} and some particular cases of them can be quantized \cite{CurlForces4}. The classical motions under these forces can be classified \cite{CurlForces1}. They also imply some topological properties of magnetic fields \cite{Kotiuga}. Therefore, understanding their structure is of paramount importance.

A decomposition of a curl force expressible by functions and their gradients was given in \cite{YavariGoriely}, namely:
\begin{itemize}
 \item Proposition 2.3 \cite{YavariGoriely}: In $\mathbb{R}^{2}$ a force $F:\mathbb{R}^{2} \rightarrow \mathbb{R}^{2}$ can be expressed as $F(x) = -V(x)\nabla U(x)$, where generalized potentials are $U,V:\mathbb{R}^{2}\rightarrow \mathbb{R}$. Moreover, $\nabla V(x)=0 \Leftrightarrow \nabla \times F =0$. More generally $V=f(U)\Leftrightarrow \nabla \times F =0$ for some function $f:\mathbb{R}\rightarrow\mathbb{R}$.
 \item Proposition 2.4 \cite{YavariGoriely}: In $\mathbb{R}^{3}$ a force $F:\mathbb{R}^{3} \rightarrow \mathbb{R}^{3}$ can be expressed as $F(x) = -V(x)\nabla U(x)-\nabla W(x)$, where generalized potentials are $U,V,W:\mathbb{R}^{3}\rightarrow \mathbb{R}$, and $\nabla V\times \nabla U \neq 0$. Moreover, $\nabla V(x)=0 \Leftrightarrow \nabla \times F =0$ and we can select $W=0$. We have $\nabla W = 0 \Leftrightarrow F\cdot \nabla\times F = 0$.
\end{itemize}
These results stem from transforming\footnote{The mappings $g:TM\rightarrow T^{*}M$ and $g^{-1}:T^{*}M\rightarrow TM$ are called musical isomorphisms \cite{Thirring, AbrahamMarsden, Frankel}.} a force $F\in TM$ into the work form $\omega = g(F,\_):TM\rightarrow \mathbb{R}$, i.e., $\omega\in T^{*}M$; applying the Darboux theorem \cite{EdelenExteriorCalculus, EdelenIsovectorMethods} then determines the local form of $\omega$, which is subsequently transformed back into a vector field on the $TM$ bundle. Due to the use of the Darboux theorem, the resulting decomposition is local in nature and not unique. This approach is therefore distinct from the global Helmholtz-Hodge decomposition \cite{deRham, Warner}, which is valid for compact manifolds. Moreover, these findings suggest that it is more natural to work with forces by expressing them as differential forms, particularly as the standard notion of a curl does not exist in higher-dimensional spaces.

The precise understanding of the decomposition of vector fields into gradient and curl-like remainders is essential in many applications that have arisen recently, namely in nonlinear elasticity \cite{YavariGoriely_Elasticity}, in the foundation of odd materials\footnote{It is interesting that D.G.B.\ Edelen influenced the thermodynamical discipline in a way that led to the odd materials formulation, as well as the mechanical/geometrical aspects of the aforementioned decomposition which we elaborate in this paper.} \cite{MarcinOstojaStarzewski_OddMaterials}, and in the emerging field of investigation of curl forces \cite{Berry_sixExamples, CurlForces1, CurlForces2, CurlForces3, CurlForces4}.

In this paper, we propose a new, algorithmic approach to decompose any force into a potential component and a generalized curl component. This is done using geometric decomposition centered at an initial point $x_{0}$. The curl component is then (non-uniquely) decomposed into terms involving generalized potentials and their gradients using the Frobenius theorem corresponding to a shifted homotopy center $x_{1}\neq x_{0}$, which avoids a problem related to the nilpotency of  the homotopy operator. Unlike the method outlined in \cite{YavariGoriely}, our proposed decomposition does not require solving PDEs to construct the general or recursive decompositions (whereas obtaining a closed connection form for the gradient-recursive case generally involves solving an integrating-factor PDE). This constructive alternative, unlike the Darboux theorem, offers a novel and practical way of working with classical potentials in any number of dimensions. To address the difficulty of defining a curl outside of three-dimensional space, we replace it with the more general notion of an antiexact differential form. The proposed method also yields a more fine-grained decomposition than the one proposed in \cite{YavariGoriely}.

The paper is organized as follows: In the next section, we summarize the key results from the theory of exterior differential forms that are needed to construct the decomposition algorithm. We then present our main result, discuss the single-center obstruction and single-center variant, and provide several examples in two and three dimensions that illustrate the algorithm application. In addition we also provide an analysis of the coalescing-centers limit and the local-to-global extension of the results. Moreover, the extension of the results to the conformal class of curl forces, where Weyl geometry is used is given. The paper ends with short Discussion and with Conclusions.

\section{Preliminaries}
In this section, we summarize results from the local behavior of exterior differential forms following \cite{EdelenExteriorCalculus, EdelenIsovectorMethods, ExteriorSystems, KyciaPoincare, KyciaPoincareCohomotopy, CopoincareHamiltonianSystem}. They will be used to construct the force decomposition in the next section.

The main object of consideration is an $n$-dimensional smooth manifold $M$ equipped with a metric tensor $g:TM\times TM\rightarrow \mathbb{R}$, since an inner product is defined in most physical problems. We will also use the exterior derivative $d:\Lambda^{k}(M)\rightarrow \Lambda^{k+1}(M)$.

On this manifold, we can associate a vector field $F\in TU$ with a $1$-form $\omega$, using the musical isomorphism $\omega = g(F,\_)$. Similarly, the gradient of a function $f:U\rightarrow \mathbb{R}$ is defined via its differential $df$ such that $df = g(\nabla f,\_)$. In the paper, the metric structure will be used only to pass from vectors to exterior differential forms.

The curl operation is defined only in three-dimensional space. This is because the cross product, and consequently the curl, is specific to three-dimensional Euclidean space, where an oriented plane (a bivector) can be uniquely represented by a vector \cite{SpacetimeAlgebra, Frankel}. This representation is difficult in curvilinear coordinates and not possible in higher-dimensional spaces. Therefore, to operate in spaces of arbitrary dimension, one must abandon the notion of the curl and switch to the more general notion of the exterior derivative acting on differential forms. We will see that the natural generalization of the curl component of a vector is an antiexact differential form, as defined below.

We also note that the Helmholtz decomposition can be constructed in vector calculus \cite{HelmholtzHigherDimensional}, where multivectors are needed.

\subsection{Geometric decomposition}
We restrict our analysis to a star-shaped subset of $M$ that lies within a single local chart, which can therefore be identified with a star-shaped subset $U\subset \mathbb{R}^{n}$. The metric tensor $g:TM\times TM\rightarrow \mathbb{R}$ can then be restricted to $U$ in a chart.

The star-shaped domain $U$ contains a central point $x_{0}\in U$ and admits a linear homotopy $F_{x_{0}}:[0,1]\times U\rightarrow U$, $F_{x_{0}}(t,x)=x_{0} + t(x-x_{0})$, which connects any point $x\in U$ with $x_{0}$. We define the linear homotopy operator (writing $H_{x_{0}}$, or simply $H$ when the center is fixed) as
\begin{equation}
H_{x_{0}}\omega = \int_{0}^{1} i_{\mathcal{K}_{x_{0}}}\omega|_{F_{x_{0}}(t,x)}t^{k-1}dt,
\end{equation}
where $\mathcal{K}_{x_{0}}=(x-x_{0})^{i}\partial_{i}$, $\omega \in \Lambda^{k}(U)$ is a $k$-differential form on $U$, and $i:\Lambda^{k}(U)\times TU\rightarrow \Lambda^{k-1}(U)$ is the insertion operator $i_{\mathcal{K}_{x_{0}}}\omega = \omega(\mathcal{K}_{x_{0}},\ldots)$. One can use a different, e.g., nonlinear, homotopy $F$ in the definition of $H$; however, the linear one makes the operator nilpotent, $H^{2}=0$. It also has the properties $dHd = d$ and $HdH=H$. We also have a homotopy invariance formula
\begin{equation}
 dH_{x_{0}} + H_{x_{0}}d = I - s_{x_{0}}^{*},
 \label{Eq_homotopyFormula}
\end{equation}
where $s_{x_{0}}^{*}$ is the pullback along the constant map $s_{x_{0}}:x_{0} \hookrightarrow U$. The map $s_{x_{0}}^{*}$ can be nonzero only on $\Lambda^{0}(U)$.
Since we will be considering $1$-forms, the pullback $s_{x_{0}}^{*}$ vanishes, and from (\ref{Eq_homotopyFormula}) it follows that $H_{x_{0}}d+dH_{x_{0}}=I$ on $\Lambda^{k}(U)$ for $k>0$. Moreover, since $(dH_{x_{0}})^{2}=dH_{x_{0}}$ and $(H_{x_{0}}d)^{2}=H_{x_{0}}d$, the operators $H_{x_{0}}d$ and $dH_{x_{0}}$ are projectors. They define a direct sum decomposition of $\Lambda(U)$ into the \textit{exact}\footnote{Since the domain is star-shaped, an exact form is also closed (i.e., an element of $\ker(d)$) by the Poincar\'{e} lemma.} vector space $\mathcal{E}^{k}(U)=\ker(d) = \operatorname{im}(dH_{x_{0}})$, and the \textit{antiexact} module over $\Lambda^{0}(U)=C^{\infty}(U)$ denoted by $\mathcal{A}^{k}_{x_{0}}(U)$, given for $k\ge 1$ by $\mathcal{A}^{k}_{x_{0}}(U)=\ker(H_{x_{0}})=\operatorname{im}(H_{x_{0}}d)=\{\omega\in \Lambda^{k}(U)\mid i_{\mathcal{K}_{x_{0}}}\omega =0\}$ (which automatically implies $\omega|_{x_{0}}=0$ for $k\ge 1$), and for $k=0$ by $\mathcal{A}^{0}_{x_{0}}(U)=\{f\in C^{\infty}(U)\mid f(x_{0})=0\}$. For $f\in \Lambda^{0}(U)$, we thus have a decomposition\footnote{If $f$ is a classical potential, then the decomposition agrees with the global gauge freedom realized by the additive group $(\mathbb{R}=\ker(d),+)$.} $f(x) = \underbrace{(f(x)-f(x_{0}))}_{\mathcal{A}} \oplus \underbrace{f(x_{0})}_{\mathcal{E}}$ as $f(x_{0})\in \ker(d)$ and $(f(x)-f(x_{0}))|_{x_{0}}=0$. In summary, these statements yield the direct sum decomposition
\begin{equation}
\Lambda^{k}(U)=\mathcal{E}^{k}(U)\oplus \mathcal{A}^{k}_{x_{0}}(U), \quad k\in\{0,\ldots,n\},
\end{equation}
which is called the \textit{geometric decomposition}. It is local and associated with a specific choice of the homotopy center $x_{0}$. A different choice of $x_{0}$ induces a different decomposition; see \cite{EdelenExteriorCalculus}. The locality of the decomposition does not pose a problem when one is interested in a specific solution valid in a neighborhood of a point. Various applications of the geometric decomposition are presented in \cite{EdelenExteriorCalculus, EdelenIsovectorMethods, ExteriorSystems, KyciaPoincare, KyciaPoincareCohomotopy, CopoincareHamiltonianSystem}.

The Poincar\'{e} lemma can be proved using the geometric decomposition as $d\omega =0 \Leftrightarrow \omega = dH\omega$, that is, if $\omega$ is closed, then it is also exact on a star-shaped subset.

We also remark that a similar construction exists on a Riemannian manifold for the codifferential operator $\delta:\Lambda^{k}(U)\rightarrow \Lambda^{k-1}(U)$, which is related to the exterior derivative via the Hodge star operator. A homotopy operator for the codifferential, known as the cohomotopy operator, also exists; see \cite{KyciaPoincareCohomotopy, KyciaSilhan}.

\subsection{Frobenius theorem}
\label{Subsec_FrobeniusTheorem}
In this subsection, we summarize the main construction related to the Frobenius theorem, following \cite{EdelenExteriorCalculus}; see also \cite{EdelenIsovectorMethods, Sternber}. The main idea is the notion of a (homogeneous) ideal $I$ of $\Lambda(U)$ that is a set closed with respect to the sum of forms and multiplication: $\alpha\wedge \beta \in I$ if $\beta\in I$ and for all $\alpha\in \Lambda(U)$. The ideal $I$ is a \textit{differential ideal} when $dI\subset I$, i.e., if the ideal is closed under the exterior derivative. In many problems, the ideal is defined by a set of $1$-forms $\{\omega_{1},\ldots,\omega_{r}\}$ and is denoted by $I(\{\omega_{1},\ldots,\omega_{r}\})$. The \textit{exterior system} of dimension $r$, $D_{r}$, is a set of $r$ linearly independent $1$-forms. It generates the ideal $I(D_{r})$. The system $D_{r}=\{\omega^{i}\}_{i=1}^{r}$ is \textit{completely integrable} if and only if there exist $r$ independent functions $\{f^{i}(x)\}_{i=1}^{r}$ such that each of the forms $\omega^{i}$ vanishes on each family of $r$-parameter $(n-r)$-dimensional surfaces $\{f^{i}(x)=c^{i}\}_{i=1}^{r}$ for $c^{i}\in \mathbb{R}$. This means, see \cite{EdelenExteriorCalculus}, that $\omega^{i}=A^{i}_{j}df^{j}$ for some nonsingular $r\times r$ matrix $A(x)$ of functions.

We have
\begin{Theorem}[Frobenius, see \cite{EdelenExteriorCalculus} Theorem 4-3.2]
An exterior differential system $D_{r}$ defined by $r$ $1$-forms $D_{r}=\{\omega^{i}\}_{i=1}^{r}$ is completely integrable if and only if the ideal $I(D_{r})$ is closed ($dI(D_{r})\subset I(D_{r})$), which is equivalent to
\begin{equation}
 d\omega^{i}=\Gamma^{i}_{j}\wedge \omega^{j},
\end{equation}
or
\begin{equation}
 \omega^{1}\wedge\ldots\wedge\omega^{r}\wedge d\omega^{i}=0, \quad i\in \{1,\ldots,r\},
\end{equation}
where $\Gamma$ is an $r\times r$ matrix of $1$-forms.
\end{Theorem}

We will consider an ideal that is generated by a single $1$-form $\Omega$ on a star-shaped domain $V\subset \mathbb{R}^{n}$ with homotopy operator $H$. We observe that in this case, the crucial role in the Frobenius theorem is played by the equation
\begin{equation}
 d\Omega = \Gamma \wedge \Omega + \Sigma,
 \label{Eq.FrobeniusEquation}
\end{equation}
where $\Gamma \in \Lambda^{1}(V)$ and $\Sigma\in \Lambda^{2}(V)$. To this equation, one must adjoin its differential consequences
\begin{equation}
 d\Sigma = \Gamma\wedge \Sigma - \Theta\wedge \Omega, \quad d\Gamma=\Theta, \quad d\Theta =0,
\end{equation}
and treat them as a system of exterior differential equations. The form $\Sigma$ is called the torsion\footnote{This is not the torsion from the Cartan moving frame method on Riemannian manifolds \cite{Sternber}. It describes an obstruction to the complete integrability of $\Omega$ with respect to the connection $\Gamma$, in analogy to the curvature $2$-form for the integrability of geodesics. It is a standard equation of fundamental theories of physics---visible when rewritten as $d\Omega-\Gamma\wedge \Omega=\Sigma$ by identifying the operator $d - \Gamma\wedge\_$ as a simplified version of the exterior covariant derivative; see \cite{KyciaSilhan}.} of the equation.
The solution of this system is given by the following
\begin{Theorem}[see \cite{EdelenExteriorCalculus}, Theorem 5-9.2]
\label{Th_FrobeniusSecomposition}
Any system of exterior equations
\begin{align}
 d\Omega &= \Gamma \wedge \Omega + \Sigma, \\
 d\Sigma &= \Gamma\wedge \Sigma - \Theta\wedge \Omega, \quad d\Gamma=\Theta, \quad d\Theta =0,
\end{align}
on a star-shaped domain $V$ with homotopy operator $H$ is equivalent to the system
\begin{align}
 d\Omega &= d\gamma \wedge \Omega + \Sigma', \label{Eq.EquivalentSystem1} \\
 d\Sigma' &= d\gamma\wedge\Sigma', \quad \gamma=H(\Gamma),
\end{align}
and has the general solution
\begin{equation}
 \Omega = e^{\gamma}(d\phi + \eta), \quad \Sigma' = e^{\gamma}d\eta,
\end{equation}
with $\Sigma' = \Sigma + (Hd\Gamma)\wedge \Omega$,
\begin{equation}
 \phi = H(e^{-\gamma}\Omega), \quad \eta = H(e^{-\gamma}\Sigma')=Hd(e^{-\gamma}\Omega),
\end{equation}
that are in $\mathcal{A}(V)$.
\end{Theorem}

Note that (\ref{Eq.EquivalentSystem1}) follows directly from substituting the geometric decomposition $\Gamma = d\gamma + Hd\Gamma$ into $d\Omega = \Gamma\wedge\Omega + \Sigma$.

When the torsion $\Sigma$ vanishes, we say that $\Omega$ is \textit{recursive} with coefficient form $\Gamma$. In this case, the equations take the form
\begin{equation}
 d\Omega = \Gamma\wedge \Omega, \quad \Theta\wedge \Omega = 0, \quad d\Gamma= \Theta,
\end{equation}
with the solution
\begin{equation}
 \Omega = e^{\gamma}(d\phi + H(\theta\wedge d\phi)),
\end{equation}
where $\theta = Hd\Gamma$ with the constraint $d\theta \wedge \Omega =0$.

When, in addition to the recursive case, $\Gamma$ is closed (hence exact on $V$, $\Gamma = dH\Gamma = d\gamma$), then $\Omega$ is called \textit{gradient recursive} and the solution is
\begin{equation}
 \Omega = e^{\gamma}d\phi, \quad \phi=H(e^{-\gamma}\Omega).
\end{equation}
We note that in this case $e^{-\gamma}\Omega \in \mathcal{E}^{1}(V)$, i.e., $e^{-\gamma}$ is an integrating factor for $\Omega$.

We will call the formula for $\Omega$ the \textit{Frobenius decomposition} for $\Omega$.

We can also relate the above decomposition to the Darboux decomposition. From (\ref{Eq.FrobeniusEquation}) by wedging with $\Omega$ we get
\begin{equation}
\Omega\wedge d\Omega = \Omega\wedge \Sigma.
\label{Eq.WedgeTorsion}
\end{equation}

The connection between the torsion $\Sigma$ and the Pfaffian integrability condition $\Omega\wedge d\Omega = 0$ requires attention as $\Sigma$ depends on the choice of $\Gamma$. From (\ref{Eq.WedgeTorsion}), vanishing torsion ($\Sigma = 0$, corresponding to the recursive and gradient-recursive cases) implies $\Omega\wedge d\Omega = 0$, meaning that $\Omega$ is completely integrable (Darboux class\footnote{We recall \cite{EdelenExteriorCalculus, ExteriorSystems} that the Darboux class of a $1$-form $\omega\in\Lambda^{1}(U)$ is defined as $K=\max_{x\in U} K(x)$, where $K(x)$ is the largest integer $m$ such that $I_{m}|_{x}\neq 0$ (with $K(x)=0$ if $\omega|_{x}=0$), and the forms $I_{m}$ are defined recursively by $I_{1}=\omega$, $I_{2}=d\omega$, $I_{2k+1}=I_{1}\wedge I_{2k}$, and $I_{2k+2}=I_{2}\wedge I_{2k}$ for $k\ge 1$. In terms of the local rank $r(x) = \max\{k\ge 0 \mid I_{2k+1}|_{x}\neq 0\}$ and $s(x) = \max\{k\ge 1 \mid I_{2k}|_{x}\neq 0\}$ (with $s(x)=0$ if $d\omega|_{x}=0$), the local Darboux class satisfies $K(x) = r(x) + s(x) + 1$ whenever $\omega|_{x}\neq 0$.} at most $2$). Conversely, if $\Omega$ is nowhere vanishing on $V$ and satisfies $\Omega\wedge d\Omega = 0$, then there exists a smooth $1$-form $\Gamma\in\Lambda^{1}(V)$ for which $\Sigma = 0$. In fact, given any smooth vector field $Z$ on $V$ normalized by $\Omega(Z) = 1$. For instance, using the metric $g$, we can define
\begin{equation}
 Z = \frac{g^{-1}(\Omega,\_)}{|\Omega|_{g}^{2}}.
 \label{Eq.DualVector}
\end{equation}
Contracting $\Omega\wedge d\Omega = 0$ with $Z$ yields $d\Omega - \Omega\wedge (i_{Z}d\Omega) = 0$, i.e.,
\begin{equation}
 d\Omega = \Gamma\wedge\Omega, \quad \Gamma = -i_{Z}d\Omega.
 \label{Eq.AlgebraicGamma}
\end{equation}
Therefore a torsion-free connection form $\Gamma$ is obtained  algebraically whenever $\Omega\wedge d\Omega = 0$ and $\Omega\neq 0$ on $V$. However, a nonvanishing torsion $\Sigma\neq 0$ for an arbitrary choice of $\Gamma$ does not imply $\Omega\wedge d\Omega \neq 0$, because $\Omega\wedge\Sigma$ may still vanish (in particular, in $\mathbb{R}^{2}$ every $3$-form vanishes identically, so $\Omega\wedge d\Omega \equiv 0$ even when $\Sigma\neq 0$, as seen in Example~\ref{Example_antiexactPart} below). On the other hand, non-integrability ($\Omega\wedge d\Omega \neq 0$, Darboux class at least $3$) guarantees that $\Sigma\neq 0$ for every choice of $\Gamma\in\Lambda^{1}(V)$.

These results do not change if $\Omega$ is a $k$-form for $k>1$; see \cite{EdelenExteriorCalculus} Section 5. One can also generalize these results to vector-valued differential forms; see \cite{EdelenExteriorCalculus, KyciaSilhan}.

\section{Results}
We now combine geometric decomposition and the Frobenius theorem to decompose the work form into simpler components. Both of these methods are already known; however, the novelty of the paper lies in their subsequent application to obtain a constructive local decomposition of arbitrary forces, in contrast to the Darboux-based approach.

We begin by considering the work $1$-form $\omega$ corresponding to a force field $F:M\rightarrow TM$, defined as $\omega = g(F,\_)$. Since our method uses the homotopy operator, we restrict our analysis to a single chart and work within a star-shaped subset $U\subset \mathbb{R}^{n}$ centered at $x_{0}\in U$.

Before stating the main theorem, we establish a fundamental obstruction showing that the Frobenius decomposition does not works yielding zero when applied to the antiexact remainder $\Omega = H_{x_{0}}d\omega$ using the same radial homotopy operator $H_{x_{0}}$.

\begin{Proposition}
\label{Prop_obstruction}
Let $U\subset\mathbb{R}^{n}$ be a star-shaped domain with center $x_{0}\in U$, and let $\Omega = H_{x_{0}}d\omega \in \mathcal{A}^{1}_{x_{0}}(U)$ be the antiexact part of a smooth $1$-form $\omega\in\Lambda^{1}(U)$. Then for every smooth function $\gamma\in C^{\infty}(U)$, the rescaled form $e^{-\gamma}\Omega$ remains in $\mathcal{A}^{1}_{x_{0}}(U)$, and we obtain
\begin{equation}
 \phi = H_{x_{0}}(e^{-\gamma}\Omega) = 0, \qquad \eta = H_{x_{0}}d(e^{-\gamma}\Omega) = e^{-\gamma}\Omega.
 \label{Eq.SameCenterCollapse}
\end{equation}
In particular, applying Theorem \ref{Th_FrobeniusSecomposition} to $\Omega$ with the same homotopy operator $H_{x_{0}}$  reproduces $\Omega = e^{\gamma}(0 + e^{-\gamma}\Omega)$ without extracting a nonzero integrable component. Therefore no nonzero $\Omega\in\mathcal{A}^{1}_{x_{0}}(U)$ can realize a smooth recursive or gradient-recursive decomposition with respect to $H_{x_{0}}$.
\end{Proposition}

\begin{Proof}
Since $\Omega = H_{x_{0}}d\omega \in \mathcal{A}^{1}_{x_{0}}(U)$, it satisfies $i_{\mathcal{K}_{x_{0}}}\Omega = 0$. Because $\mathcal{A}^{1}_{x_{0}}(U)$ is a $C^{\infty}(U)$-module \cite{EdelenExteriorCalculus}, multiplication by the smooth scalar $e^{-\gamma}$ preserves this property, giving
\begin{equation}
 i_{\mathcal{K}_{x_{0}}}(e^{-\gamma}\Omega) = e^{-\gamma}i_{\mathcal{K}_{x_{0}}}\Omega = 0.
\end{equation}
Hence $e^{-\gamma}\Omega \in \mathcal{A}^{1}_{x_{0}}(U) = \ker(H_{x_{0}})$, which immediately gives $\phi = H_{x_{0}}(e^{-\gamma}\Omega) = 0$. From the homotopy invariance formula (\ref{Eq_homotopyFormula}) on $1$-forms, it follows that $\eta = H_{x_{0}}d(e^{-\gamma}\Omega) = (I - dH_{x_{0}})(e^{-\gamma}\Omega) = e^{-\gamma}\Omega$. Substituting $\phi = 0$ into the recursive or gradient-recursive expressions of Theorem \ref{Th_FrobeniusSecomposition} forces $\Omega = 0$.
\end{Proof}

Proposition \ref{Prop_obstruction} demonstrates that to extract an integrable component $e^{\gamma}d\phi$ from a nonzero antiexact remainder $\Omega = H_{x_{0}}d\omega$ within a smooth framework, one must use a different homotopy center for the Frobenius theorem, i.e., a point $x_{1}\in U$ where $\Omega(x_{1})\neq 0$ ($x_{1}\neq x_{0}$, since $\Omega(x_{0}) = 0$). This two-center construction is a base of the following main theorem.

\begin{Theorem}
\label{Th_decomposition}
For a smooth work $1$-form $\omega=g(F, \_)$ on a star-shaped subset $U\subset \mathbb{R}^{n}$ with initial center $x_{0}\in U$, we have the geometric decomposition
\begin{equation}
 \omega = df + \Omega,
\end{equation}
where $f = H_{x_{0}}\omega\in \Lambda^{0}(U)$ and $\Omega=H_{x_{0}}d\omega \in \mathcal{A}^{1}_{x_{0}}(U)$.

If $\Omega\not\equiv 0$, we choose a second center $x_{1}\in U$ such that $\Omega(x_{1})\neq 0$ and a star-shaped open neighborhood $V\subset U$ of $x_{1}$ on which $\Omega$ is nowhere vanishing. Applying Theorem \ref{Th_FrobeniusSecomposition} to $\Omega|_{V}$ with the shifted radial homotopy operator $H_{x_{1}}$, we can further decompose $\Omega$ on $V$ according to $d\Omega = \Gamma\wedge\Omega + \Sigma$, where $\Gamma\in\Lambda^{1}(V)$, $\Sigma\in\Lambda^{2}(V)$:
\begin{itemize}
 \item General case:
 \begin{equation}
 \Omega = e^{\gamma}(d\phi + \eta), \quad \Sigma' = e^{\gamma}d\eta,
 \label{Eq.MainGeneral1}
\end{equation}
with $\gamma = H_{x_{1}}\Gamma$, $\Sigma' = \Sigma + (H_{x_{1}}d\Gamma)\wedge \Omega$, and
\begin{equation}
 \phi = H_{x_{1}}(e^{-\gamma}\Omega), \quad \eta = H_{x_{1}}(e^{-\gamma}\Sigma')=H_{x_{1}}d(e^{-\gamma}\Omega).
 \label{Eq.MainGeneral2}
\end{equation}
Moreover, as $d\phi|_{x_{1}} = \Omega|_{x_{1}} \neq 0$, so $\phi$ is non-constant on every neighborhood of $x_{1}$.
\item Recursive case ($\Omega\wedge d\Omega = 0$, $\Sigma=0$, realized by $\Gamma = -i_{Z}d\Omega$ by equations (\ref{Eq.DualVector})-(\ref{Eq.AlgebraicGamma})):
\begin{equation}
 \Omega = e^{\gamma}(d\phi + H_{x_{1}}(\theta\wedge d\phi)),
 \label{Eq.MainRecursive}
\end{equation}
where $\theta = H_{x_{1}}d\Gamma$, with the always satisfied constraint $d\theta \wedge \Omega =0$.
\item \textit{Gradient recursive case} ($\Sigma=0$, $d\Gamma = 0$, i.e., $\Gamma = dH_{x_{1}}\Gamma = d\gamma$):
\begin{equation}
 \Omega = e^{\gamma}d\phi, \quad \phi=H_{x_{1}}(e^{-\gamma}\Omega).
 \label{Eq.MainGradRecursive}
\end{equation}
\end{itemize}
The functions $f,\gamma,\phi$ can be called generalized potentials. When $\Omega \equiv 0$, the second step is omitted and $\omega = df$ reduces to a standard conservative force with potential $f$.
\end{Theorem}

\begin{Proof}
We use the notation and definitions from the previous section. The proof begins with the geometric decomposition of $\omega$ on $U$ relative to $x_{0}$:
\begin{equation}
 \omega = dH_{x_{0}}\omega + H_{x_{0}}d\omega.
\end{equation}
We identify in the exact (gradient) part the potential $f=H_{x_{0}}\omega \in \Lambda^{0}(U)$. The second term, $\Omega = H_{x_{0}}d\omega\in \mathcal{A}^{1}_{x_{0}}(U)$, is the antiexact (curl) part.

If $\Omega =0$, then $\omega=df$ is exact, and the force $F$ is a conservative gradient.

In the case where $\Omega\not\equiv 0$, there exists $x_{1}\in U$ with $\Omega(x_{1})\neq 0$ (and $x_{1}\neq x_{0}$ because $\Omega(x_{0}) = 0$). By continuity, $\Omega$ is nowhere vanishing on an open ball $V\subset U$ around $x_{1}$, which is star-shaped with respect to $x_{1}$. Applying Theorem \ref{Th_FrobeniusSecomposition} to $\Omega|_{V}$ using $H_{x_{1}}$ gives equations (\ref{Eq.MainGeneral1})-(\ref{Eq.MainGradRecursive}). Moreover, as $\gamma = H_{x_{1}}\Gamma$ and $\eta = H_{x_{1}}d(e^{-\gamma}\Omega)$ satisfy $\gamma(x_{1}) = 0$ and $\eta|_{x_{1}} = 0$, evaluating $\Omega = e^{\gamma}(d\phi + \eta)$ at $x_{1}$ gives $d\phi|_{x_{1}} = \Omega|_{x_{1}} \neq 0$, showing that the integrable component $d\phi$ near $x_{1}$ is nontrivial.

In the recursive case, differentiating $d\Omega = \Gamma\wedge\Omega$ gives $0 = d\Gamma\wedge\Omega = d\theta\wedge\Omega$ automatically.
\end{Proof}

In three dimensions, the extraction of $df = dH_{x_{0}}\omega$ can reduce the Darboux class of the curl force, giving a two-gradient representation $\omega = df + e^{\gamma}d\phi$ (matching Proposition 2.4 of \cite{YavariGoriely}) even when $\omega\wedge d\omega \neq 0$. We have the following
\begin{Proposition}
\label{Prop_3D_homogeneous}
Let $U\subset\mathbb{R}^{3}$ be a star-shaped domain with center $x_{0}\in U$, and let $\omega\in\Lambda^{1}(U)$ be a smooth $1$-form whose exterior derivative $d\omega$ has coefficients that are functions homogeneous of degree $m-1\ge 0$ in the shifted coordinates $y = x - x_{0}$ (equivalently, $\Omega = H_{x_{0}}d\omega$ has components homogeneous of degree $m\ge 1$ in $y = x - x_{0}$). Then $\Omega\wedge d\Omega \equiv 0$ on $U$, even if $\omega\wedge d\omega \neq 0$.
\end{Proposition}
\begin{Proof}
Expanding $\Omega = \Omega_{i}(y)dy^{i}$ with each $\Omega_{i}$ homogeneous of degree $m$ in $y = x - x_{0}$, homogeneity (Euler's theorem) gives $\mathcal{L}_{\mathcal{K}_{x_{0}}}(\Omega_{i}) = m\Omega_{i}$ and $\mathcal{L}_{\mathcal{K}_{x_{0}}}(dy^{i}) = dy^{i}$, hence $\mathcal{L}_{\mathcal{K}_{x_{0}}}\Omega = (m+1)\Omega$. By Cartan's formula and the antiexactness condition $i_{\mathcal{K}_{x_{0}}}\Omega = 0$, we get
\begin{equation}
 (m+1)\Omega = \mathcal{L}_{\mathcal{K}_{x_{0}}}\Omega = i_{\mathcal{K}_{x_{0}}}d\Omega + d(i_{\mathcal{K}_{x_{0}}}\Omega) = i_{\mathcal{K}_{x_{0}}}d\Omega.
\end{equation}
Contracting the $3$-form $\Omega\wedge d\Omega$ with $\mathcal{K}_{x_{0}}$ therefore gives
\begin{equation}
 i_{\mathcal{K}_{x_{0}}}(\Omega\wedge d\Omega) = (i_{\mathcal{K}_{x_{0}}}\Omega)\wedge d\Omega - \Omega\wedge(i_{\mathcal{K}_{x_{0}}}d\Omega) = -(m+1)\Omega\wedge\Omega = 0.
\end{equation}
Since $\Omega\wedge d\Omega$ is a top-degree form on $\mathbb{R}^{3}$ and $\mathcal{K}_{x_{0}}(x) = x - x_{0} \neq 0$ for all $x\neq x_{0}$, it follows that $\Omega\wedge d\Omega = 0$ on $U\setminus\{x_{0}\}$, and by continuity on all of $U$.
\end{Proof}

We now comment on a single-center case.
\begin{Remark}
\label{Remark_single_center}
We can consider the single-center case, where we do not separate the exact part $dH_{x_{0}}\omega$ first, and if $\omega$ is nowhere vanishing on a star-shaped domain $U$ centered at $x_{0}$ (so $\omega(x_{0})\neq 0$, excluding equilibrium points). Then Theorem \ref{Th_FrobeniusSecomposition} can be applied to $\omega$ using the single operator $H_{x_{0}}$. This yields $\omega = e^{\gamma}(d\phi + \eta)$ on $U$ with $d\phi|_{x_{0}} = \omega|_{x_{0}} \neq 0$. Setting $\Gamma = 0$ recovers the additive geometric decomposition $\omega = df + \Omega$. In the gradient-recursive case (which is always possible locally in $\mathbb{R}^{2}$ since $\omega\wedge d\omega \equiv 0$), it produces the minimal two-potential representation $\omega = e^{\gamma}d\phi$ of Proposition 2.3 in \cite{YavariGoriely}. However, in $\mathbb{R}^{3}$ a single-center case cannot achieve the class reduction of Proposition \ref{Prop_3D_homogeneous} when $\omega\wedge d\omega \neq 0$, nor can it be centered at a force equilibrium $F(x_{0}) = 0$.
\end{Remark}

\begin{Remark}
The form $\Omega \in \mathcal{A}^{1}_{x_{0}}(U)$ is a generalization of the curl part of the force.
\end{Remark}

\begin{Remark}
For historical reasons, the definition of potential used in physics is $-f$. However, in this presentation, it is more natural to use this function without the minus sign.
\end{Remark}

Transforming the results to the tangent space we obtain the following:
\begin{Corollary}
A vector field $F:U\rightarrow TU$ on a star-shaped $U\subset \mathbb{R}^{n}$ with center $x_{0}\in U$ can be decomposed as
\begin{equation}
 F = \nabla f + X,
\end{equation}
where, if $X\not\equiv 0$, then on any star-shaped open subset $V\subset U$ with center $x_{1}\in V$ on which $X$ is nowhere vanishing, we have
\begin{itemize}
 \item \textit{General case:} $X = e^{\gamma}\nabla \phi + Y$, where $Y=e^{\gamma}g^{-1}(\eta,\_)$.
 \item \textit{Recursive case:} $X=e^{\gamma}\nabla \phi + e^{\gamma}Y$, where $Y = g^{-1}(H_{x_{1}}(\theta\wedge d\phi),\_)$.
 \item \textit{Gradient recursive case:} $X = e^{\gamma}\nabla \phi$.
\end{itemize}
\end{Corollary}

\begin{Remark}
Note that for the general and recursive cases there is a vector $Y$ in the decomposition which cannot be expressed as a gradient of a function.
\end{Remark}

The above corollary mixes covariant and contravariant objects, since the decomposition is performed on differential forms and then transformed back to the tangent space. The natural way to work with such a decomposition is at the level of differential forms.

From Theorem \ref{Th_decomposition}, we can infer the algorithm:
\begin{Algorithm}
\label{Algorithm}
\textbf{Input:} A smooth force $F$ and a metric tensor $g$.
\begin{enumerate}
 \item Select the initial homotopy center $x_{0}$ and define a star-shaped subset $U\subset \mathbb{R}^{n}$ for $x_{0}$.
 \item Compute the work form $\omega = g(F,\_)$.
 \item Extract the potential $f=H_{x_{0}}\omega$.
 \item Identify the generalized curl component $\Omega=H_{x_{0}}d\omega$.
 \item If $\Omega \equiv 0$, return the conservative decomposition $\omega = df$.
 \item Else ($\Omega\not\equiv 0$) select a shifted center $x_{1}\in U$ with $\Omega(x_{1})\neq 0$ and a star-shaped open neighborhood $V\subset U$ of $x_{1}$ on which $\Omega$ is nowhere vanishing, and test complete integrability via $\Omega\wedge d\Omega$ on $V$:
 \begin{enumerate}
  \item If $\Omega\wedge d\Omega = 0$ set $Z = g^{-1}(\Omega,\_)/|\Omega|_{g}^{2}$ and $\Gamma = -i_{Z}d\Omega$ (yielding $\Sigma = 0$), and compute $\gamma = H_{x_{1}}\Gamma$ and $\phi = H_{x_{1}}(e^{-\gamma}\Omega)$. If $d\Gamma = 0$, return the gradient-recursive form $\Omega = e^{\gamma}d\phi$. Otherwise compute $\theta = H_{x_{1}}d\Gamma$ and return the recursive form $\Omega = e^{\gamma}(d\phi + H_{x_{1}}(\theta\wedge d\phi))$ (Optionally looking for a gauge transform $\tilde{\Gamma} = \Gamma + h\Omega$ that gives closed $\tilde{\Gamma}$, i.e., $d\tilde{\Gamma}=0$, to eliminate $\theta$, where an arbitrary smooth function $h$ is used. This preserves $\tilde{\Gamma}\wedge\Omega = \Gamma\wedge\Omega=d\Omega$).
  \item If $\Omega\wedge d\Omega \neq 0$ (or for any chosen smooth gauge $\Gamma\in\Lambda^{1}(V)$) compute $\gamma = H_{x_{1}}\Gamma$, $\phi = H_{x_{1}}(e^{-\gamma}\Omega)$, and $\eta = H_{x_{1}}d(e^{-\gamma}\Omega)$, and return the general decomposition $\Omega = e^{\gamma}(d\phi + \eta)$.
 \end{enumerate}
\end{enumerate}
\end{Algorithm}

\begin{Remark}
\label{Remark_nonarbitrarness}
In the general case, the selection of $\Gamma$ and $\Sigma$ for $\Omega\neq 0$ can be simplified. One may directly select a smooth positive factor $e^{-\gamma}$ on $V$ (with $\gamma(x_{1})=0$), which fixes the closed connection form $\Gamma=d\gamma$ ($\Theta = 0$, so $\Sigma' = \Sigma$), and then compute $\phi = H_{x_{1}}(e^{-\gamma}\Omega)$ and $\eta = H_{x_{1}}d(e^{-\gamma}\Omega)$ without explicitly evaluating $\Sigma$. By contrast, while both the general decomposition and the recursive decomposition by (\ref{Eq.AlgebraicGamma}) are completely PDE-free, converting a non-closed torsion-free connection $\Gamma$ ($d\Gamma\neq 0$) into a closed integrating gauge $\tilde{\Gamma} = \Gamma + h\Omega = d\tilde{\gamma}$ for the gradient-recursive case in general requires solving the first-order PDe for integrating-factor $d\tilde{\gamma}\wedge\Omega = d\Omega$.
\end{Remark}

We summarize the comparison between the Darboux decomposition and the proposed algorithm in Table \ref{Tab.Comparison}.
\begin{table}[htbp]
\centering
\small
\begin{tabular}{|p{2cm} | p{4cm} | p{5cm}|}
\hline
\textbf{Feature} & \textbf{Darboux approach \cite{YavariGoriely}} & \textbf{Algorithm \ref{Algorithm}} \\
\hline
\textbf{Integrability} & Based on $\omega\wedge d\omega$ & Based on $\Omega\wedge d\Omega$ (existence of $\Gamma$ with $\Sigma = 0$); $3$ cases \\ \hline
\textbf{Computation} & Solving PDEs & Integrations ($H_{x_{0}}, H_{x_{1}}$) for general and recursive cases; PDE only if converting non-closed $\Gamma$ to gradient-recursive \\  \hline
\textbf{Uniqueness} & Non-unique & Non-unique (gauge choice, homotopy centers $x_{0}, x_{1}$) \\  \hline
\textbf{Dimensionality} & 2D/3D & Arbitrary $n$-dimensions \\  \hline
\textbf{Components} & Potential and Curl & Potential and General/Recursive/ Gradient-Recursive Curl (1 or 2 components) \\ \hline
\end{tabular}
\caption{Comparison of the Darboux approach \cite{YavariGoriely} and Algorithm \ref{Algorithm}.}
\label{Tab.Comparison}
\end{table}

\subsection{Non-uniqueness as a gauge freedom}
\label{Subsec_GaugeFreedom}
Once the initial homotopy center $x_{0}$ is fixed, the antiexact part $\Omega$ is a geometric object that has a fixed transformation under the change of the center, see \cite{EdelenExteriorCalculus}.

On the subdomain $V$ centered at $x_{1}$, the choice of $\Gamma$ and $\Sigma$ is arbitrary as long as they fulfill $d\Omega = \Gamma\wedge \Omega + \Sigma$. The selection of $\Gamma$, and consequently $\gamma = H_{x_{1}}\Gamma$, can be seen as a gauge choice. The factor $e^{-\gamma}$ can be chosen to simplify the decomposition. Different choices of this factor balance the complexity between the integrable part $e^{\gamma}d\phi$ and the path-dependent part $e^{\gamma}\eta$.
We summarize this in the following
\begin{Theorem}
\label{Th_GaugeScale}
For $\Omega \neq 0$ on a star-shaped subdomain $V\subset U$ with center $x_{1}\in V$, under the change of gauge
\begin{equation}
 \Gamma \rightarrow \Gamma + d\lambda
\end{equation}
by $\lambda \in C^{\infty}(V)$, the curvature $\Theta = d\Gamma$ is invariant, while $\gamma = H_{x_{1}}\Gamma$ and $\Sigma$ transform as
\begin{align}
 \gamma & \rightarrow \gamma + \lambda - \lambda(x_{1}), \\
 \Sigma &\rightarrow \Sigma - d\lambda\wedge\Omega.
\end{align}
In particular, if $\lambda\in\mathcal{A}^{0}_{x_{1}}(V)$ (i.e., $\lambda(x_{1}) = 0$), then $\gamma \rightarrow \gamma + \lambda$.
\end{Theorem}
\begin{Proof}
Let $\hat{\Gamma} = \Gamma + d\lambda$ for an arbitrary $\Gamma\in\Lambda^{1}(V)$ satisfying $d\Omega = \Gamma\wedge\Omega + \Sigma$ and any $\lambda\in C^{\infty}(V)$. Then $d\hat{\Gamma} = d\Gamma = \Theta$. Applying the homotopy operator $H_{x_{1}}$ and using homotopy invariance formula (\ref{Eq_homotopyFormula}) on $0$-forms ($H_{x_{1}}d\lambda = \lambda - \lambda(x_{1})$) yields $\hat{\gamma} = H_{x_{1}}\hat{\Gamma} = \gamma + \lambda - \lambda(x_{1})$.

Since $d\Omega = \Gamma \wedge\Omega + \Sigma$ must be preserved by the change of gauge, we have $d\Omega = (\hat{\Gamma}-d\lambda)\wedge \Omega + \Sigma$. Comparing this with $d\Omega = \hat{\Gamma}\wedge\Omega + \hat{\Sigma}$ gives $\hat{\Sigma} = \Sigma - d\lambda\wedge \Omega$, as required.
\end{Proof}
The specific choice of $\Gamma$ can be motivated by a physical situation, e.g., by the preservation of symmetry. Another choice can be aimed at 'minimizing' the torsion $\Sigma$ to decrease the level of non-integrability of $\Omega$.

\subsection{Examples}

We now provide several examples of the Algorithm \ref{Algorithm} application.

\begin{Example}
The simplest example is an exact work form. Consider $\mathbb{R}^{2}$ with the origin as its homotopy center and with the standard Euclidean metric. Take the work form
\begin{equation}
 \omega = xdx = d\left( \frac{1}{2} x^{2}\right).
\end{equation}
The potential is $f= \frac{1}{2} x^{2}$.

We can verify that this agrees with the geometric decomposition. We have
\begin{equation}
 dH\omega = d\left(\frac{1}{2}x^{2}\right) = \omega,
\end{equation}
\begin{equation}
 Hd\omega = 0 \quad \Rightarrow \quad \omega = dH\omega,
\end{equation}
as required.

The decomposition $\omega=dH\omega + Hd\omega$ is therefore satisfied, with the antiexact part being zero.
\end{Example}

\begin{Example}
\label{Example_antiexactPart}
In this example, we consider an antiexact work form on $\mathbb{R}^{2}$ with the initial homotopy center at the origin $x_{0}=(0,0)$ and with the standard Euclidean metric. We select the form
\begin{equation}
 \omega = xdy - ydx.
\end{equation}
We find that $d\omega = 2dx\wedge dy$, so the form is not exact. Applying the geometric decomposition at $x_{0}=(0,0)$ yields
\begin{equation}
 dH_{x_{0}}\omega = d\left(\frac{1}{2}(xy-yx)\right)=0,
\end{equation}
\begin{equation}
 \Omega = H_{x_{0}}d\omega = \omega.
\end{equation}
Therefore, $\omega$ is purely antiexact at the origin.

Following Algorithm \ref{Algorithm}, we restrict to the convex (hence star-shaped) half-plane $V = \{(x,y)\in\mathbb{R}^{2} \mid x > 0\}$ with shifted center $x_{1}=(1,0)$, where $\Omega|_{(1,0)} = dy \neq 0$. The linear homotopy on $V$ is $F_{x_{1}}(t;x,y) = (1 + t(x-1), ty)$ with $\mathcal{K}_{x_{1}} = (x-1)\partial_{x} + y\partial_{y}$.

We now compare three smooth gauge choices on $V$. We start from the trivial gauge (general case, $\Sigma\neq 0$). Choosing $\Gamma = 0$ gives $\gamma = 0$ and $\Sigma = \Sigma' = 2dx\wedge dy$. Integration yields
 \begin{equation}
  \phi = H_{x_{1}}\Omega = \int_{0}^{1} \left((1 + t(x-1))y - ty(x-1)\right)dt = y,
 \end{equation}
 \begin{equation}
  \eta = H_{x_{1}}\Sigma' = H_{x_{1}}(2dx\wedge dy) = (x-1)dy - ydx,
 \end{equation}
 giving $\Omega = d\phi + \eta = dy + ((x-1)dy - ydx)$.

 Note that $\Sigma\neq 0$ even though $\Omega\wedge d\Omega \equiv 0$ in $\mathbb{R}^{2}$, showing that nonvanishing torsion for a non-adapted gauge does not imply non-integrability.

 The next example is closed integrating gauge (gradient recursive, $\Sigma = 0$). By selecting the smooth closed $1$-form $\Gamma = \frac{2}{x}dx = d\ln(x^{2})$ on $V$ we get $\Gamma\wedge\Omega = 2dx\wedge dy = d\Omega$, so $\Sigma = 0$ and $\Theta = 0$. Applying $H_{x_{1}}$ yields
 \begin{equation}
  \gamma = H_{x_{1}}\Gamma = \ln(x^{2}), \quad \phi = H_{x_{1}}(x^{-2}\Omega) = H_{x_{1}}\left(d\frac{y}{x}\right) = \frac{y}{x}, \quad \eta = 0,
 \end{equation}
 giving $\Omega = x^{2}d(y/x)$ on $V$.

 The final case is an algebraic gauge (\ref{Eq.AlgebraicGamma}). We obtain $Z = \frac{-y\partial_{x} + x\partial_{y}}{x^{2}+y^{2}}$, with normalization $\Omega(Z)=1$. Equation (\ref{Eq.AlgebraicGamma}) gives $\Gamma = -i_{Z}d\Omega = \frac{2(xdx + ydy)}{x^{2}+y^{2}} = d\ln(x^{2}+y^{2})$, which is smooth and closed on $V$. Then $\gamma = H_{x_{1}}\Gamma = \ln(x^{2}+y^{2})$ and $\phi = H_{x_{1}}\left(\frac{xdy - ydx}{x^{2}+y^{2}}\right) = \arctan(y/x)$, yielding $\Omega = (x^{2}+y^{2})d\arctan(y/x)$.

Comparing these three smooth gauges on $V$ illustrates both the non-uniqueness of the Frobenius decomposition and the possibility of elimination of torsion by selecting a gauge.
\end{Example}

\begin{Example}
\label{Example_mixed}
Consider a work form on $\mathbb{R}^{2}$, which we take to be star-shaped with the origin $x_{0}=(0,0)$ as its initial homotopy center
\begin{equation}
 \omega = xdx+y^{2}dx.
\end{equation}
The first term $xdx$ is exact ($d(xdx)=0$) and is equal to $d(\frac{1}{2}x^{2})$. However, the second term is not exact, since $d(y^{2}dx)=2ydy\wedge dx \neq 0$.

First, we decompose $\omega$ at $x_{0}=(0,0)$ into its exact and antiexact parts
\begin{equation}
 dH_{x_{0}}\omega = d\left(\frac{1}{2}x^{2}+\frac{1}{3}xy^{2} \right)= xdx+\frac{2}{3}xydy+\frac{1}{3}y^{2}dx,
\end{equation}
\begin{equation}
 \Omega = H_{x_{0}}d\omega = \omega-dH_{x_{0}}\omega = H_{x_{0}}(2ydy\wedge dx)=\frac{2}{3}(y^{2}dx-xydy).
\end{equation}
Note that both terms of the original form $\omega$ contribute to the exact part. This shows that even if a component of a form is not exact, it can still contribute to the exact part of the total form. The potential for the exact term is
\begin{equation}
 f = \frac{1}{2}x^{2}+\frac{1}{3}xy^{2}.
\end{equation}

Next, we apply the Frobenius decomposition to $\Omega$ on the convex open first quadrant $V = \{(x,y)\in\mathbb{R}^{2} \mid x > 0, y > 0\}$ with the center $x_{1} = (1,1)$, where $\Omega(1,1) = \frac{2}{3}(dx - dy) \neq 0$. Choosing the smooth closed connection $1$-form $\Gamma = \frac{dx}{x} + \frac{2dy}{y} = d\ln(xy^{2})$ on $V$, we have $\Gamma\wedge\Omega = -2ydx\wedge dy = d\Omega$, so $\Sigma = 0$ and $\Theta = 0$. Applying $H_{x_{1}}$ gives
\begin{equation}
 \gamma = H_{x_{1}}\Gamma = \ln(xy^{2}), \quad \phi = H_{x_{1}}(e^{-\gamma}\Omega) = H_{x_{1}}\left(\frac{2}{3}\left(\frac{dx}{x} - \frac{dy}{y}\right)\right) = \frac{2}{3}\ln\left(\frac{x}{y}\right),
\end{equation}
with $\eta = 0$. Thus $\Omega = xy^{2}d\left(\frac{2}{3}\ln(x/y)\right)$ is a smooth gradient-recursive decomposition on $V$, with both $\gamma$ and $\phi$ normalized to vanish at $x_{1}=(1,1)$.

For comparison, the single-center direct variant of Remark \ref{Remark_single_center} can be applied directly to $\omega = (x+y^{2})dx$ on the convex half-plane $U' = \{(x,y)\in\mathbb{R}^{2}\mid x > 0\}$ with center $x_{0}' = (1,0)$, where $x + y^{2} > 0$. Taking the closed gauge $\Gamma_{1} = d\ln(x+y^{2})$ gives $\gamma = \ln(x+y^{2})$ and $\phi = x - 1$, yielding $\omega = (x+y^{2})dx$. By contrast, choosing $Z = (x+y^{2})^{-1}\partial_{x}$ in (\ref{Eq.AlgebraicGamma}) produces $\Gamma_{2} = -i_{Z}d\omega = \frac{2y}{x+y^{2}}dy$, which satisfies $\Sigma = 0$ but has $d\Gamma_{2} = \frac{-2y}{(x+y^{2})^{2}}dx\wedge dy \neq 0$.

This illustrates the difference between the recursive and gradient-recursive cases. We have $\Gamma_1 - \Gamma_2 = \frac{dx}{x+y^2} = \frac{1}{(x+y^2)^2}\omega$. Here the gauge  choice for making $\Gamma_{1}$ closed is $h=\frac{1}{(x+y^2)^2}$.
\end{Example}

\begin{Example}
\label{Example_3D}
To illustrate higher dimensional case, we consider two work forms on $\mathbb{R}^{3}$ with the standard Euclidean metric and initial homotopy center $x_{0} = (0,0,0)$.

First, we illustrate the reduction for a homogeneous curl. We take $\omega = dz + xdy$. Here $d\omega = dx\wedge dy$ and $\omega\wedge d\omega = dx\wedge dy\wedge dz \neq 0$ everywhere, so $\omega$ itself is of Darboux class $3$. Following the Algorithm \ref{Algorithm} in the first step at $x_{0}=(0,0,0)$ we get
 \begin{equation}
  f = H_{x_{0}}\omega = z + \frac{1}{2}xy, \qquad \Omega = H_{x_{0}}d\omega = \frac{1}{2}(xdy - ydx).
 \end{equation}
According to the Proposition \ref{Prop_3D_homogeneous} ($m=1$), $\Omega\wedge d\Omega \equiv 0$. Choosing $V = \{(x,y,z)\in\mathbb{R}^{3}\mid x > 0\}$ with the new center $x_{1} = (1,0,0)$ and closed gauge $\Gamma = \frac{2}{x}dx$ gives $\gamma = \ln(x^{2})$ and $\phi = \frac{y}{2x}$, yielding the two-gradient decomposition $\omega = d\left(z + \frac{1}{2}xy\right) + x^{2}d\left(\frac{y}{2x}\right)$.

The second case is an irreducible general case (Darboux class $3$ for $\Omega$). We take $\omega = xdy + y^{2}dz$, whose exterior derivative $d\omega = dx\wedge dy + 2ydy\wedge dz$ is non-homogeneous. Applying the first step of Algorithm \ref{Algorithm} at $x_{0}=(0,0,0)$ gives
 \begin{equation}
  f = H_{x_{0}}\omega = \frac{1}{2}xy + \frac{1}{3}y^{2}z, \quad \Omega = H_{x_{0}}d\omega = \frac{1}{2}(xdy - ydx) + \frac{2}{3}y(ydz - zdy).
 \end{equation}
 This gives
 \begin{equation}
  \Omega\wedge d\Omega = -\frac{1}{3}y^{2}dx\wedge dy\wedge dz \neq 0 \quad (y\neq 0),
 \end{equation}
 so $\Sigma\neq 0$ for every smooth connection form $\Gamma$ on any domain away from $y=0$. Restricting to the convex half-space $V = \{(x,y,z)\in\mathbb{R}^{3}\mid y > 0\}$ with the new center $x_{1} = (0,1,0)$, where $\Omega(0,1,0) = -\frac{1}{2}dx + \frac{2}{3}dz \neq 0$, and selecting the trivial gauge $\Gamma = 0$ ($\gamma = 0$), integration with $F_{x_{1}}(t;x,y,z) = (tx, 1 + t(y-1), tz)$ produces
 \begin{eqnarray}
  \phi = H_{x_{1}}\Omega = -\frac{1}{2}x + \frac{1}{3}(y+1)z, \\ \eta = H_{x_{1}}d\Omega = \frac{1}{2}\big(xdy - (y-1)dx\big) + \frac{1}{3}(2y+1)\big((y-1)dz - zdy\big),
 \end{eqnarray}
yielding $\Omega = d\phi + \eta$ with a non-removable (path-dependent) curl $\eta$.
\end{Example}

We can notice that in Example \ref{Example_mixed} we can observe that $\gamma$ and $\phi$ are singular when $x_{1}= (1,1)$ merges with $x_{0}=(0,0)$. This is general feature of the method and is described in the following subsection.

\subsection{Merging homotopy centers and renormalization}

A natural question regarding Theorem \ref{Th_decomposition} is how the two-center decomposition behaves when the shifted homotopy center $x_{1}$ is pushed towards $x_{0}$ after the decomposition has been computed. We show that the limiting behavior divides into two regimes depending on whether the chosen connection $1$-form $\Gamma$ remains smooth at $x_{0}$ or develops a singularity there.

\textbf{$\Gamma$ is smooth:} We assume first that $\Gamma$ is smooth on a convex neighborhood $U_{0}\subset U$ of $x_{0}$, as occurs in the general case (i.e., with the trivial gauge $\Gamma = 0$). As $x_{1}\rightarrow x_{0}$ within $U_{0}$, the linear homotopy $F_{x_{1}}(t,x) = x_{1} + t(x - x_{1})$ and the Euler vector field $\mathcal{K}_{x_{1}} = (x - x_{1})^{i}\partial_{i}$ converge  on compact subsets to $F_{x_{0}}$ and $\mathcal{K}_{x_{0}}$. Consequently, the potentials and remainder satisfy $\gamma_{x_{1}} = H_{x_{1}}\Gamma \rightarrow \gamma_{x_{0}} = H_{x_{0}}\Gamma$, $\phi_{x_{1}} = H_{x_{1}}(e^{-\gamma_{x_{1}}}\Omega) \rightarrow 0$, and $\eta_{x_{1}} = H_{x_{1}}d(e^{-\gamma_{x_{1}}}\Omega) \rightarrow e^{-\gamma_{x_{0}}}\Omega$ continuously. This gives the results of Proposition \ref{Prop_obstruction}. The limits also can be inferred from the change of center formula for homotopy operators in \cite{EdelenExteriorCalculus}.

The rate of this limiting process can be computed. We set $\alpha = e^{-\gamma_{x_{1}}}\Omega \in \mathcal{A}^{1}_{x_{0}}(U_{0})$ and evaluate the antiexactness identity $(y - x_{0})^{i}\alpha_{i}(y) = 0$ (resulting from $i_{\mathcal{K}_{x_{0}}}\alpha=0$) along the homotopy ray $y(t) = x_{1} + t(x - x_{1})$. Since $y(t) - x_{0} = (x_{1} - x_{0}) + t(x - x_{1})$, we obtain $t(x - x_{1})^{i}\alpha_{i}(y(t)) = -(x_{1} - x_{0})^{i}\alpha_{i}(y(t))$, and hence
\begin{equation}
 \phi_{x_{1}}(x) = \int_{0}^{1} (x - x_{1})^{i}\alpha_{i}(y(t))dt = -\int_{0}^{1} \frac{(x_{1} - x_{0})^{i}\alpha_{i}(y(t))}{t}dt.
 \label{Eq.CoalescingSmooth}
\end{equation}
Since $(x_{1} - x_{0})^{i}\alpha_{i}(x_{1}) = 0$ at $t=0$, the quotient in (\ref{Eq.CoalescingSmooth}), by Hadamard's lemma, extends smoothly to $t=0$ and scales linearly with the separation vector $(x_{1}-x_{0})$. This yields $\phi_{x_{1}}(x) = O(|x_{1} - x_{0}|)$ uniformly on compact subsets of $U_{0}$. As an example we consider the case from Example \ref{Example_antiexactPart} with the trivial gauge $\Gamma = 0$ and a general shifted center $x_{1} = (a,b)$. Integration gives $\phi_{x_{1}}(x,y) = ay - bx$ and $\eta_{x_{1}} = (x - a)dy - (y - b)dx$. They satisfy $\phi_{x_{1}} \rightarrow 0$ and $\eta_{x_{1}} \rightarrow \Omega$ as $(a,b) \rightarrow (0,0)$ as required.

\textbf{Torsion-free ($\Sigma=0$) case:} The situation changes in the torsion-free (recursive and gradient-recursive) cases. As observed in the above examples this case yields singularities. We have the following
\begin{Proposition}
 No nonzero antiexact $1$-form $\Omega \in \mathcal{A}^{1}_{x_{0}}(U)$ can satisfy $d\Omega = \Gamma\wedge\Omega$ for a connection $1$-form $\Gamma$ that is smooth on a star-shaped neighborhood $U_{0}\subset U$ of $x_{0}$.
\end{Proposition}
\begin{Proof}
 We assume that $\Gamma$ is smooth and decompose it at $x_{0}$ as $\Gamma = d\gamma + \theta$, where $\gamma = H_{x_{0}}\Gamma \in C^{\infty}(U_{0})$ and $\theta = H_{x_{0}}d\Gamma \in \mathcal{A}^{1}_{x_{0}}(U_{0})$. As before, we define $\alpha = e^{-\gamma}\Omega \in \mathcal{A}^{1}_{x_{0}}(U_{0})$. Then we get
 \begin{equation}
  d\alpha = e^{-\gamma}(d\Omega - d\gamma\wedge\Omega) = \theta\wedge\alpha.
 \end{equation}
 Since both $\theta$ and $\alpha$ are antiexact with the center at $x_{0}$, we have $i_{\mathcal{K}_{x_{0}}}(\theta\wedge\alpha) = 0$, and therefore, $H_{x_{0}}(\theta\wedge\alpha) = 0$. Applying the homotopy invariance formula (\ref{Eq_homotopyFormula}), we get $\alpha = dH_{x_{0}}\alpha + H_{x_{0}}d\alpha$, which gives $\alpha = H_{x_{0}}(\theta\wedge\alpha) = 0$, and therefore $\Omega \equiv 0$ on $U_{0}$ which contradicts assumption that $\Omega$ is nonzero. As a result we get that every torsion-free gauge $\Gamma$ for a nonzero $\Omega \in \mathcal{A}^{1}_{x_{0}}(U)$ is therefore necessarily singular at $x_{0}$.
\end{Proof}
\begin{Corollary}
 Non-vanishing torsion allows the connection $\Gamma$ to be smooth at $x_{0}$.
\end{Corollary}
For general $\Omega$ the estimation of leading order singularities for $\gamma$ when centers merge is complicated. We therefore focus on the case limit in the gradient-recursive case and for simplicity we assume that $\Omega \in \mathcal{A}^{1}_{x_{0}}(U)$ has components that are homogeneous functions of degree $m \ge 1$ in the shifted coordinates $y = x - x_{0}$, and that $d\Omega = \Gamma\wedge\Omega$ holds with a smooth closed connection form $\Gamma$ ($d\Gamma = 0$) on a convex cone $V\subset U$ with $x_{0} \in \partial V$ on which $\Omega$ is nowhere vanishing. Then, as in the proof of Proposition \ref{Prop_3D_homogeneous}, homogeneity gives
\begin{equation}
 i_{\mathcal{K}_{x_{0}}}d\Omega = (m+1)\Omega.
\end{equation}
Contracting $d\Omega = \Gamma\wedge\Omega$ with $\mathcal{K}_{x_{0}}$ and using $i_{\mathcal{K}_{x_{0}}}\Omega = 0$ then gives $(m+1)\Omega = (i_{\mathcal{K}_{x_{0}}}\Gamma)\Omega$. As a result we have $i_{\mathcal{K}_{x_{0}}}\Gamma = m+1$ everywhere on $V$.

\begin{figure}
\centering
 \includegraphics[width=0.8\textwidth]{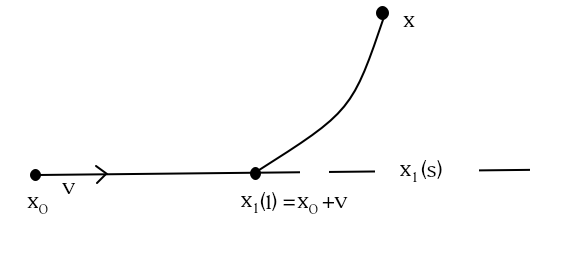}
 \caption{The geometry of integration.}
 \label{Fig.IntegrationGeometry}
\end{figure}

Now we can isolate singularity ($s\rightarrow 0^{+}$) by moving to the second center toward $x_{0}$ along the ray $x_{1}(s) = x_{0} + sv \in V$, with $v\neq 0$, and integrating the closed form $\Gamma$ from $x_{1}(s)$ to $x_{0} + v$ and then from $x_{0} + v$ to $x$, see Fig. \ref{Fig.IntegrationGeometry}.

For the segment from $x_0 + sv$, $0<s \leq 1$, to the point $x_0 + v$, we use the path with tangent vector
\begin{eqnarray}
 c(\tau) = x_0 + \tau v \quad \text{for} \quad \tau \in [s, 1], \\
 c'(\tau) = v = \frac{1}{\tau}(\tau v) = \frac{1}{\tau}\mathcal{K}_{x_0}(c(\tau)).
\end{eqnarray}
This yields
\begin{equation}
 \Gamma(c'(\tau)) = \frac{1}{\tau}\Gamma(\mathcal{K}_{x_0}) = \frac{m+1}{\tau}.
\end{equation}

Summing up these two contributions we get
\begin{equation}
 \gamma_{x_{1}(s)}(x) = \int_{s}^{1} \frac{m+1}{\tau}d\tau + \tilde{\gamma}_{v}(x) = \tilde{\gamma}_{v}(x) - (m+1)\ln s, \quad e^{\gamma_{x_{1}(s)}(x)} = s^{-(m+1)}e^{\tilde{\gamma}_{v}(x)},
 \label{Eq.CoalescingGamma}
\end{equation}
where $\tilde{\gamma}_{v}(x) = \int_{x_{0}+v}^{x}\Gamma$ is smooth on $V$ and independent of $s$.

Similarly, the closed (path-independent) $1$-form $e^{-\tilde{\gamma}_{v}}\Omega$ is annihilated by $i_{\mathcal{K}_{x_{0}}}$, so its line integral along the radial segment from $x_{0}+sv$ to $x_{0}+v$ vanishes identically. We get therefore
\begin{equation}
 \phi_{x_{1}(s)}(x) = \int_{x_{1}(s)}^{x} e^{-\gamma_{x_{1}(s)}}\Omega = \int_{x_{1}(s)}^{x} s^{m+1} e^{-\tilde{\gamma}_v} \Omega  = s^{m+1}\tilde{\phi}_{v}(x),
 \label{Eq.CoalescingPhi}
\end{equation}
where
\begin{equation}
 \tilde{\phi}_{v}(x) = \int_{x_{0}+v}^{x} e^{-\tilde{\gamma}_{v}}\Omega.
\end{equation}
As $s\rightarrow 0^{+}$, the scaling factor $e^{\gamma_{x_{1}(s)}}$ diverges as $s^{-(m+1)}$ while the potential $\phi_{x_{1}(s)}$ vanishes as $s^{m+1}$, and the two powers of $s$ cancel out so that $\Omega = e^{\gamma_{x_{1}(s)}}d\phi_{x_{1}(s)}$ remains exact for every $s > 0$.

This cancellation is visible in Examples \ref{Example_antiexactPart} and \ref{Example_mixed}: for $\Omega = xdy - ydx$ ($m=1$, $\Gamma = \frac{2}{x}dx$) with $x_{1} = (a,b)$ in $\{x>0\}$, one obtains $e^{\gamma_{x_{1}}} = x^{2}/a^{2}$ and $\phi_{x_{1}} = a^{2}(y/x - b/a)$, whereas for $\Omega = \frac{2}{3}(y^{2}dx - xydy)$ ($m=2$, $\Gamma = d\ln(xy^{2})$) with $x_{1} = (a,b)$ in $\{x>0, y>0\}$, one finds $e^{\gamma_{x_{1}}} = xy^{2}/(ab^{2})$ and $\phi_{x_{1}} = \frac{2}{3}ab^{2}\ln(bx/(ay))$.

Equations (\ref{Eq.CoalescingGamma}) and (\ref{Eq.CoalescingPhi}) show that the divergence as $x_{1}(s)\rightarrow x_{0}$ is an artifact of the homotopy center normalization $\gamma_{x_{1}(s)}(x_{1}(s)) = 0$ imposed by $H_{x_{1}(s)}$. We can absorb the diverging constant using the gauge shift $\tilde{\gamma}_{v} = \gamma_{x_{1}(s)} + (m+1)\ln s$ at the same time compensating by the rescaling $\tilde{\phi}_{v} = s^{-(m+1)}\phi_{x_{1}(s)}$. This removes all dependence on $s$ and yields finite, $x_{1}$-independent renormalized potentials $\tilde{\gamma}_{v}$ and  $\tilde{\phi}_{v}$ on $V$ that satisfy $\Omega = e^{\tilde{\gamma}_{v}}d\tilde{\phi}_{v}$. As $i_{\mathcal{K}_{x_{0}}}d\tilde{\phi}_{v} = 0$, the renormalized potential $\tilde{\phi}_{v}$ is homogeneous of degree zero in $x - x_{0}$ (purely angular and therefore singular at $x_{0}$, see, e.g., Example \ref{Example_antiexactPart} or \ref{Example_mixed}), while $i_{\mathcal{K}_{x_{0}}}d\tilde{\gamma}_{v} = m+1$ implies that $e^{\tilde{\gamma}_{v}}$ is homogeneous of degree $m+1$ and vanishes at $x_{0}$ (so $\tilde{\gamma}_{v}\rightarrow -\infty$ as $x\rightarrow x_{0}$). In this sense, keeping the second homotopy center separated from $x_{0}$ serves as a 'geometric regularization' of the singular renormalized potentials at the origin.

\subsection{From local to global}
In applications, typically, the local decomposition is the most useful. However, in geometric considerations, the connection of the above results to the global properties of the underlying manifold is essential.

Consider a good cover\footnote{An open cover where all open sets are contractible along with their finite intersections.} $\{U_{\alpha}\}_{\alpha}$ of a manifold $M$. Denote $U_{\alpha\beta} = U_{\alpha}\cap U_{\beta}$. Using the local decomposition of a $1$-form $\omega\in \Lambda^{1}(M)$ in $U_{\alpha}$ and $U_{\beta}$ as
\begin{align}
 \omega|_{U_{\alpha}} &= df_{\alpha}+\Omega_{\alpha}, \\
 \omega|_{U_{\beta}} &= df_{\beta}+\Omega_{\beta},
\end{align}
and assuming $U_{\alpha\beta}\neq \emptyset$, on the overlap we have
\begin{equation}
 d(f_{\alpha}-f_{\beta}) = \Omega_{\beta}-\Omega_{\alpha}.
 \label{Eq.ConsistencyMatching}
\end{equation}
We can consider two cases:\\
\textbf{Case 1}: When local radial antiexact parts $\Omega_{\alpha} = H_{x_{\alpha}}d\omega$ depend on the chart centers $x_{\alpha}\in U_{\alpha}$, they do not generally glue into a single global form unless $d\omega = 0$ (in which case $\Omega_{\alpha} \equiv 0$). More generally, if there exists a global $1$-form $\Omega\in\Lambda^{1}(M)$ satisfying $d\Omega = d\omega$ so that we may take $\Omega_{\alpha} = \Omega|_{U_{\alpha}}$, then $\omega - \Omega$ is closed and (\ref{Eq.ConsistencyMatching}) reduces to
\begin{equation}
 d(f_{\alpha}-f_{\beta}) = 0.
\end{equation}
Since $U_{\alpha\beta}$ is contractible, we have
\begin{equation}
f_{\alpha}-f_{\beta} = c_{\alpha\beta} \in \mathbb{R}.
\label{Eq.fa-fb=cab}
\end{equation}
Considering the cocycle condition on $U_{\alpha\beta\gamma} = U_{\alpha}\cap U_{\beta} \cap U_{\gamma}\neq \emptyset$, we have, using (\ref{Eq.fa-fb=cab}),
\begin{equation}
c_{\alpha\beta}+c_{\beta\gamma}+c_{\gamma\alpha} = 0.
\end{equation}
Therefore the coefficients $\{c_{\alpha\beta}\}$ define a $1$-cocycle in \v{C}ech cohomology $\check{H}^{1}(M, \mathbb{R})\cong H^{1}_{\mathrm{dR}}(M)$ representing the de Rham class $[\omega - \Omega]$. If this cohomology class is non-trivial, then we cannot define a global potential function $f\in C^{\infty}(M)$ such that $f_{\alpha}=f|_{U_{\alpha}}$.\\
\textbf{Case 2}: In the case when $\Omega_{\beta}-\Omega_{\alpha} \neq 0$, exterior differentiating (\ref{Eq.ConsistencyMatching}) yields the consistency condition
\begin{equation}
 d\Omega_{\alpha} = d\Omega_{\beta}
\end{equation}
on $U_{\alpha\beta}$ (consistent with $d\Omega_{\alpha} = d\omega|_{U_{\alpha}}$). In conclusion to this case, we obtain
\begin{Corollary}
Even if antiexact parts of the decomposition do not agree on an overlap, their exterior derivatives are equal.
\end{Corollary}

\subsection{Amplitude modulation of curl forces and Weyl geometry}
One can also rewrite equation (\ref{Eq.FrobeniusEquation}) for the antiexact part $\Omega$ of the work form as
\begin{equation}
 \dcov \Omega = \Sigma,
 \label{Eq.CovariantConstantOmega}
\end{equation}
where
\begin{equation}
 \dcov = d - \Gamma\wedge\_.
 \label{Eq.Lichnerowicz}
\end{equation}
The operator $\dcov$ bears a resemblance to the exterior covariant derivative on the trivial line bundle (or fiber set \cite{KyciaSilhan, EdelenExteriorCalculus}) $V\times \mathbb{R}\rightarrow V$. In the general problem of parallel transport on a bundle, one provides the connection form $\Gamma$ and an external force $\Sigma$ to compute $\Omega$, as was presented in \cite{KyciaSilhan, EdelenExteriorCalculus}. However, for (\ref{Eq.FrobeniusEquation}), we obtain the inverse problem: determine $\Gamma$ and $\Sigma$ for a fixed $\Omega$. This is a non-unique problem. Equation (\ref{Eq.CovariantConstantOmega}) also suggests that the antiexact force is a solution of a non-uniform parallel transport problem on a bundle with connection form $\Gamma$ and non-uniformity term $\Sigma$. As we infer in \cite{KyciaSilhan}, the antiexact part $\Omega$ represents an external influence on the system, and equation (\ref{Eq.CovariantConstantOmega}) can be interpreted as an evolution of this external agent, seen through the force $\Omega$ with which it influences the system. Moreover, the curvature for this external system is given by\footnote{In general, the curvature of a connection form $\Gamma$ is given by $\Theta=d\Gamma + \Gamma\wedge \Gamma$. However, for an $\mathbb{R}$-valued $1$-form the term $\Gamma\wedge \Gamma$ vanishes.} $\Theta = d\Gamma$.

We can extend this picture to the Weyl geometry, see \cite{WeylManifolds} for basic introduction and \cite{CapSlovakWeyl2003, CapSlovakWeyl2022} for advanced one, when instead of a single curl force $\Omega$ we focus on the conformal class consisting of $e^{\lambda}\Omega$ forces, for different smooth rescalings $\lambda$, that represents amplitude modulation of curl force. Up to now the work $1$-form
\begin{equation}
 \omega = df + \Omega
\end{equation}
was treated as a single fixed physical force, and the gauge freedom $\Gamma \rightarrow \Gamma + d\lambda$ redistributed complexity between the scaling factor $e^{\gamma}$ and the torsion $\Sigma$ while keeping $\Omega$ unchanged, as in Theorem \ref{Th_GaugeScale}. We now shift perspective from a single curl force to a family of forces generated by spatially modulating the amplitude of the antiexact component $\Omega$.

This modulation can be made on curl/antiexact part in isolation due to the fact that for the geometric decomposition $\Lambda^{1}(U) = \mathcal{E}^{1}(U) \oplus \mathcal{A}^{1}_{x_{0}}(U)$, the $\mathcal{A}^{1}_{x_{0}}(U)$ is a $C^{\infty}(U)$-module \cite{EdelenExteriorCalculus}, and $\mathcal{E}^{1}(U)$ is only a real vector space. Consequently, multiplying the antiexact part by any smooth positive function leaves the conservative part completely unaffected and preserves antiexactness (curl property). We therefore have the following
\begin{Proposition}
\label{Prop_modulation_invariance}
Let $\omega = df + \Omega$ be the geometric decomposition of a smooth work $1$-form on a star-shaped domain $U\subset\mathbb{R}^{n}$ centered at $x_{0}\in U$, with $f = H_{x_{0}}\omega$ and $\Omega = H_{x_{0}}d\omega \in \mathcal{A}^{1}_{x_{0}}(U)$. For any smooth function $\lambda\in C^{\infty}(U)$, define the amplitude-modulated work form
\begin{equation}
 \hat{\omega} = df + \hat{\Omega}, \qquad \hat{\Omega} = e^{\lambda}\Omega.
 \label{Eq.ModulatedWorkForm}
\end{equation}
Then the geometric decomposition of $\hat{\omega}$ with respect to $x_{0}$ recovers the exact same conservative part $dH_{x_{0}}\hat{\omega} = df$ and the modulated antiexact part $H_{x_{0}}d\hat{\omega} = e^{\lambda}\Omega$.
\end{Proposition}
\begin{Proof}
Since $\Omega \in \mathcal{A}^{1}_{x_{0}}(U) = \ker(H_{x_{0}})$ satisfies $i_{\mathcal{K}_{x_{0}}}\Omega = 0$, the rescaled form $\hat{\Omega} = e^{\lambda}\Omega$ fulfils $i_{\mathcal{K}_{x_{0}}}\hat{\Omega} = e^{\lambda}i_{\mathcal{K}_{x_{0}}}\Omega = 0$, so $\hat{\Omega} \in \mathcal{A}^{1}_{x_{0}}(U)$ and $H_{x_{0}}\hat{\Omega} = 0$. Applying $dH_{x_{0}}$ to (\ref{Eq.ModulatedWorkForm}) yields
\begin{equation}
 dH_{x_{0}}\hat{\omega} = dH_{x_{0}}(df) + dH_{x_{0}}\hat{\Omega} = df,
\end{equation}
and therefore $H_{x_{0}}d\hat{\omega} = \hat{\omega} - dH_{x_{0}}\hat{\omega} = e^{\lambda}\Omega$.
\end{Proof}

Proposition \ref{Prop_modulation_invariance} allows us to fix the conservative background $df$ and analyze the work form as an equivalence class of forces with respect to curl amplitude modulation
\begin{equation}
 [\omega]_{f} = \big\{ df + e^{\lambda}\Omega \;\big|\; \lambda \in C^{\infty}(U) \big\},
\end{equation}
or equivalently, to study the conformal class
\begin{equation}
 [\Omega] = \{e^{\lambda}\Omega \mid \lambda \in C^{\infty}(V)\}
\end{equation}
on a star-shaped neighborhood $V\subset U$ of $x_{1}$ where $\Omega$ is nowhere vanishing.

Although $\hat{\Omega} = e^{\lambda}\Omega$ remains antiexact, its exterior derivative (the generalized curl) acquires an additional term
\begin{equation}
 d\hat{\Omega} = d(e^{\lambda}\Omega) = e^{\lambda}(d\Omega + d\lambda \wedge \Omega).
 \label{Eq.ModulatedCurl}
\end{equation}
In the covariant formulation $\dcov \Omega = \Sigma$ with $\dcov = d - \Gamma\wedge\_$, this extra term is absorbed by a simultaneous shift of the connection $1$-form $\Gamma$ under the active (gauge) transformation
\begin{equation}
 \hat{\Omega} = e^{\lambda}\Omega, \qquad \hat{\Gamma} = \Gamma + d\lambda, \qquad \hat{\Sigma} = e^{\lambda}\Sigma.
 \label{Eq.WeylGaugeTransform}
\end{equation}
Then the covariant exterior derivative transforms covariantly
\begin{equation}
 \hat{\dcov}\hat{\Omega} = (d - \hat{\Gamma}\wedge\_)(e^{\lambda}\Omega) = e^{\lambda}\big(d\Omega + d\lambda\wedge\Omega - (\Gamma + d\lambda)\wedge\Omega\big) = e^{\lambda}\dcov\Omega = \hat{\Sigma},
\end{equation}
while the curvature $\Theta = d\Gamma$ remains invariant ($\hat{\Theta} = d\hat{\Gamma} = \Theta$). Furthermore, the normalized $1$-form $e^{-\gamma}\Omega$ rescales only by a constant factor $e^{\lambda(x_{1})}$ (and is strictly invariant whenever $\lambda(x_{1}) = 0$, since $\hat{\gamma} = H_{x_{1}}\hat{\Gamma} = \gamma + \lambda - \lambda(x_{1})$).

This establishes a direct formal correspondence with Weyl geometry\footnote{The relation is with connection structure and its action on conformal class of differential forms, and is not related to metric structure as in full Weyl geometry.} on the trivial real line bundle $V\times\mathbb{R}\rightarrow V$. Here, $\Gamma$ plays the role of a Weyl (scale) connection $1$-form governing local rescalings of the work form, $\dcov = d - \Gamma\wedge\_$ is the Weyl exterior derivative acting on weight-$1$ forms, $\Theta = d\Gamma$ is the Weyl (length) curvature, and $\Sigma$ measures the failure of $\Omega$ to be Weyl-closed. The three cases of Theorem \ref{Th_decomposition} then can be reformulated in a geometric terms over the class $[\Omega]$.

\textbf{Gradient-recursive case ($\Sigma = 0$, $\Theta = 0$)} represents flat (integrable) Weyl structure. The covariance constant equation $\dcov \Omega = 0$ states that $\Omega$ is Weyl-closed with respect to a flat Weyl connection ($d\Gamma = 0$). Parallel transport of scale around any contractible loop is trivial, so $\Gamma = d\gamma$ is exact on $V$. Performing the Weyl transformation (\ref{Eq.WeylGaugeTransform}) with $\lambda = -\gamma$ gauges the connection away completely ($\hat{\Gamma} = 0$) and reduces $\hat{\Omega} = e^{-\gamma}\Omega = d\phi$ to an ordinary exact form. Geometrically, the entire class $[\Omega]$ is locally conformally exact, that means, every force in the class is a conservative gradient $d\phi$ measured in a position-dependent scale $e^{\gamma}$.

\textbf{Recursive case ($\Sigma = 0$, $\Theta \neq 0$)} represents curved Weyl structure. Here $\Omega$ is still Weyl-closed ($\dcov\Omega = 0$), which implies $\Omega \wedge d\Omega = 0$ and therefore preserves complete Frobenius integrability on $[\Omega]$. However, the selected Weyl connection has non-vanishing curvature $\Theta = d\Gamma \neq 0$, meaning that scale transport around closed loops is path-dependent (analogous to the second clock effect in Weyl geometry\footnote{Synchronized clocks in Weyl spacetime traverse different paths which eventually intersects. Then the time on each clock can be different (primary effect) and the time scales of a 'tick' can be also different (secondary effect). We focus on pure Weyl geometry and not the gravity with Weyl scaling, for difference see \cite{SecondCloclEffect_Experiment, NoSecondaryClockEffectWeylGeometryGravity}.} \cite{WeylGeometry, SecondCloclEffect_Experiment, NoSecondaryClockEffectWeylGeometryGravity}). Consequently, $\Gamma$ cannot be eliminated by a scalar rescaling (\ref{Eq.WeylGaugeTransform}). Its antiexact curvature potential $\theta = H_{x_{1}}\Theta$ does not vanish in the decomposition
\begin{equation}
\Omega = e^{\gamma}(d\phi + H_{x_{1}}(\theta\wedge d\phi)).
\end{equation}
Note that since $\Omega\wedge d\Omega = 0$ is invariant under (\ref{Eq.WeylGaugeTransform}), a flat compatible Weyl connection $\tilde{\Gamma} = \Gamma + h\Omega$ (see Algorithm \ref{Algorithm} point 6.a), where $h$ is a smooth function, with $d\tilde{\Gamma} = 0$ always exists locally, although, finding $h$ generally requires solving an integrating-factor PDE.

\textbf{General case ($\Sigma \neq 0$)} represents non-Weyl-closed (torsional) structure. When $\Omega\wedge d\Omega \neq 0$, no choice of Weyl connection $\Gamma$ can make $\Omega$ Weyl-closed, see discussion at the end of Section \ref{Subsec_FrobeniusTheorem}. Then the torsion in $\Sigma = \dcov\Omega$ is non-removable. Applying (\ref{Eq.WeylGaugeTransform}), the $3$-form $\Omega\wedge\Sigma = \Omega\wedge d\Omega$ transforms as $\hat{\Omega}\wedge\hat{\Sigma} = e^{2\lambda}(\Omega\wedge\Sigma)$, showing that the Pfaffian non-integrability ($\Omega\wedge d\Omega\neq 0$) is a conformal invariant of the class $[\Omega]$. Note that in $\mathbb{R}^{3}$, it is proportional\footnote{The proportionality function comes from possible volume form for Riemannian case.} to the helicity density $X\cdot(\nabla\times X)$ of the curl field. The remainder $\eta = H_{x_{1}}d(e^{-\gamma}\Omega)$ describes this obstruction to conformal exactness. This obstruction cannot be reduced, as from $\Omega\wedge d\Omega \neq 0$ results that there is no $\gamma$ such that $d(e^{-\gamma}\Omega) = 0$. Then we have $d\eta = d H_{x_1} d(e^{-\gamma}\Omega) = d(e^{-\gamma}\Omega) \neq 0$, so there is no integrating factor $e^{-\gamma}$ for $\Omega$.

\begin{Remark}
The relation between the passive gauge freedom of Section \ref{Subsec_GaugeFreedom}, Theorem \ref{Th_GaugeScale}, and the active Weyl symmetry (\ref{Eq.WeylGaugeTransform}) can be seen when considering a single two-parameter gauge transformation. For any two independent smooth functions $\mu, \lambda \in C^{\infty}(V)$, consider the simultaneous amplitude modulation of $\Omega$ and gauge shift of $\Gamma$
\begin{equation}
 \hat{\Omega} = e^{\mu}\Omega, \quad \hat{\Gamma} = \Gamma + d\lambda.
 \label{Eq.GeneralTwoParamTransform}
\end{equation}
While the curvature $\Theta = d\Gamma$ remains invariant and $\hat{\gamma} = H_{x_{1}}\hat{\Gamma} = \gamma + \lambda - \lambda(x_{1})$, computing $\hat{\dcov}\hat{\Omega} = (d - \hat{\Gamma}\wedge\_)(e^{\mu}\Omega)$ shows that the torsion $\Sigma = \dcov\Omega$ transforms as
\begin{equation}
 \hat{\Sigma} = e^{\mu}\big(\Sigma + d(\mu - \lambda)\wedge\Omega\big).
 \label{Eq.GeneralTorsionTransform}
\end{equation}
Two special cases follow immediately from (\ref{Eq.GeneralTwoParamTransform})-(\ref{Eq.GeneralTorsionTransform}):
\begin{enumerate}
 \item Passive gauge transformation ($\mu = 0$, fixed $\Omega$) is realized by setting $\mu = 0$. It leaves the curl force unchanged ($\hat{\Omega} = \Omega$) while shifting the connection $\hat{\Gamma} = \Gamma + d\lambda$ and transforming the torsion inhomogeneously as $\hat{\Sigma} = \Sigma - d\lambda\wedge\Omega$. This recovers Theorem \ref{Th_GaugeScale} of Section \ref{Subsec_GaugeFreedom}.
 \item Active Weyl gauge transformation ($\mu = \lambda$) is realized by 'synchronizing' the amplitude modulation with the connection shift. It eliminates the inhomogeneous term $d(\mu - \lambda)\wedge\Omega$ in (\ref{Eq.GeneralTorsionTransform}), yielding $\hat{\Omega} = e^{\lambda}\Omega$, $\hat{\Gamma} = \Gamma + d\lambda$, and the covariant torsion law $\hat{\Sigma} = e^{\lambda}\Sigma$ of (\ref{Eq.WeylGaugeTransform}). This preserves the torsion-free condition $\Sigma = 0$ and acts as a genuine Weyl symmetry on the conformal class $[\Omega]$.
\end{enumerate}
\end{Remark}

We can provide mechanical interpretation of this approach. Transforming $[\omega]_{f}$ back to the tangent bundle via $g^{-1}$ yields the family of force fields
\begin{equation}
 \hat{F} = \nabla f + e^{\lambda}X,
\end{equation}
where the conservative part $\nabla f$ is fixed. As multiplying $\Omega$ by $e^{\lambda} > 0$ preserves its kernel $\ker(\Omega) = \{v \in TV \mid \Omega(v) = 0\}$, every force in the class shares the same equilibrium/critical points set of the curl part ($\Omega = 0$), the same direction of the curl force at every point, and the same zero-work directions along which the non-conservative component performs no virtual work ($\Omega(\dot{x}) = 0$). In the torsion-free cases ($\Sigma = 0$), all forces in $[\omega]_{f}$ share the same foliation of integrable hypersurfaces $\phi = \text{const}$ orthogonal to $X$. The factor $e^{\lambda}$ modulates only the spatial amplitude or coupling profile of the curl force. Typical example is varying the spatial intensity profile of an optical vortex \cite{CurlForces2} without altering its geometric circulation structure, while leaving the underlying conservative potential $f$ unaffected. This can be seen as inducing (effective) Weyl structure on the work $1$-form of a particle moving in a modulated optical field.

This approach has also thermodynamical interpretation in the Carath\'{e}odory formulation of thermodynamics \cite{Frankel, YavariGoriely}, where an infinitesimal heat exchange is represented by a $1$-form $\delta Q$ on the (thermodynamical) state space. Identifying the antiexact part $\Omega$ with $\delta Q$ (and $df$ with an exact internal-energy-like differential), the conformal class $[\Omega]$ corresponds to rescaling the heat form by a positive state function while preserving the adiabatic distribution given by $\ker(\delta Q)$. In the gradient-recursive case ($\Sigma = 0, \Theta = 0$), $\Omega = e^{\gamma}d\phi$ directly mirrors $\delta Q = TdS$, where $e^{\gamma}$ plays the role of the absolute temperature $T$ (an inverse integrating factor) and $\phi$ is the empirical entropy $S$. A Weyl transformation (\ref{Eq.WeylGaugeTransform}) rescales the temperature $T \rightarrow e^{\lambda}T$ while leaving the adiabatic hypersurfaces $S = \text{const}$ invariant. The recursive case ($\Sigma = 0, \Theta \neq 0$) corresponds to an integrable adiabatic foliation described with a non-closed (path-dependent) temperature gauge. Finally, the general case ($\Sigma \neq 0$, $\Omega\wedge d\Omega \neq 0$) represents a violation of Carath\'{e}odory's accessibility condition, where no integrating factor (temperature) exists globally or locally on $V$ to reduce the non-conservative exchange to a single entropy potential. We note that Weyl geometry has also been applied in continuum thermo-mechanics to model thermal strains and metrical defects in solids \cite{WeylGeometryInThermoelasticity}, where the Weyl structure acts on the material metric rather than on the work $1$-form itself.

\section{Discussion}
The exact part of the work form is associated with the conservative force, a classical result in mechanics. The antiexact part, as noted in \cite{KyciaSilhan}, can be thought of as the influence of an external agent acting on the system.

Further decomposition of curl part using the Frobenius theorem (evaluated on a subdomain $V$ around a shifted center $x_{1}$ where $\Omega(x_{1})\neq 0$) extracts the integrable part, $e^{\gamma}d\phi$, which can be interpreted in terms of inaccessibility, as in the Carath\'{e}odory interpretation of thermodynamics, see \cite{Frankel, YavariGoriely, YavariGoriely_Elasticity, MarcinOstojaStarzewski_OddMaterials}. The term $e^{\gamma}d\phi$ represents the force that is aligned along the (integrable) submanifolds, while $e^{\gamma}\eta$ does not define such a geometric structure.

In the gradient-recursive case, this is the only term comprising the antiexact part $\Omega$. It can be interpreted as a standard potential-driven force derived from $\phi$, but where the system coupling to this potential is modulated by a position-dependent scaling factor $e^{\gamma}$. Physically, it could model phenomena such as a charged particle moving in a field whose effective coupling varies in a specific way.

For the general and recursive cases (but not for the gradient-recursive case), the term $e^{\gamma}\eta$ is also present. It represents a more significant obstruction to integrability, as it cannot be written as the exterior derivative of a function multiplied by another function. This term represents the 'path-dependent core' of the force, which cannot be eliminated by a simple scaling factor. It can be present in driven systems where the external agent input is not derivable from any potential, scaled or otherwise. Note that this term is present in the general case (when torsion $\Sigma\neq 0$) or in the recursive case (when torsion vanishes, however, when $\Gamma$ has an antiexact part). It therefore measures the torsion and the antiexactness of $\Gamma$.

These terms open a new way to split the system dynamics among these different factors.

The approach from the paper can also be applied to non-conservative forces, e.g., in hyperelasticity \cite{YavariGoriely_Elasticity}, where the work form is an essential construction. The approach presented in the paper can lead to decomposing the work form into smaller components that can be further analyzed. Here $\Omega$ represents the non-potential part, and $\eta$ represents the fundamental obstruction to integrability (or potentiality) and enables us to quantify it.

\section{Conclusions}
A local decomposition for classical forces in a space of arbitrary dimension is proposed. This method employs geometric decomposition on a star-shaped subset of a manifold centered at $x_{0}$, distinguishing the exact part, associated with the gradient of a potential, from the antiexact part, which generalizes the curl component of the force. After establishing that applying the Frobenius theorem with the same radial homotopy operator $H_{x_{0}}$ collapses identically, we resolve the antiexact part using the (non-unique) Frobenius decomposition relative to a shifted center $x_{1}\neq x_{0}$ into an integrable component and a remainder that cannot be represented as the exterior derivative of a function.

The proposed decomposition is constructive and can be applied locally on any smooth manifold. An algorithm for its implementation has also been presented. A key advantage of this method is that it does not require solving partial differential equations (PDEs) for the general and recursive decompositions - solving PDE for the gradient-recursive case is pointed out. We also establish a three-dimensional class-reduction property for homogeneous curls, analyze the coalescing-centers limit and its renormalization, and note a single-center variant. Moreover, we extend results to conformal class of curl forces and show the connections with Weyl geometry along with simple applications. Discussion of the possible meaning of each term is also provided.

\section*{Acknowledgments}
The author declares no conflict of interest.\\
No data is attached to the presented research.\\
The work was partially funded by the EU Horizon Europe MSCA grant No 101183077.
The author acknowledges the support of the COST CA24122 and CA22123 actions.
The author would like to thank Michael V.\ Berry and Pragya Shukla for the enlightening discussion on the nature of curl forces.


\end{document}